\documentclass[pra,showpacs,floatfix]{revtex4}
\usepackage{amsmath}
\usepackage{amssymb}
\usepackage{graphicx}
\usepackage{graphicx}

\def\ih{h\kern-0.6em\char"16\kern-0.1em}

\begin{document}

\title{Two routes to the one-dimensional discrete nonpolynomial Schr\"{o}%
dinger equation}
\author{G. Gligori\'{c}$^{1}$, A. Maluckov$^{2}$, L. Salasnich$^{3}$, B. A.
Malomed$^{4}$ and Lj. Had\v{z}ievski$^{1}$}
\affiliation{$^1$ Vin\v ca Institute of Nuclear Sciences, P.O. Box 522,11001 Belgrade,
Serbia\\
$^2$ Faculty of Sciences and Mathematics, University of Ni\v s, P.O. Box
224, 18001 Ni\v s, Serbia \\
$^3$ CNR-INFM and CNISM, Department of Physics "Galileo Galilei", University
of Padua, Via Marzolo 8, 35131 Padua, Italy\\
$^4$ Department of Physical Electronics, School of Electrical Engineering,
Faculty of Engineering, Tel Aviv University, Tel Aviv 69978, Israel}

\begin{abstract}
The Bose-Einstein condensate (BEC), confined in a combination of the
cigar-shaped trap and axial optical lattice, is studied in the framework of
two models described by two versions of the one-dimensional (1D) discrete
nonpolynomial Schr\"{o}dinger equation (NPSE). Both models are derived from
the three-dimensional Gross-Pitaevskii equation (3D GPE). To produce ``model
1" (which was derived in recent works), the 3D GPE is first reduced to the
1D continual NPSE, which is subsequently discretized. ``Model 2", that was
not considered before, is derived by first discretizing the 3D GPE, which is
followed by the reduction of the dimension. The two models seem very
different; in particular, model 1 is represented by a single discrete
equation for the 1D wave function, while model 2 includes an additional
equation for the transverse width. Nevertheless, numerical analyses show
similar behaviors of fundamental unstaggered solitons in both systems, as
concerns their existence region and stability limits. Both models admit the
collapse of the localized modes, reproducing the fundamental property of the
self-attractive BEC confined in tight traps. Thus, we conclude that the
fundamental properties of discrete solitons predicted for the strongly
trapped self-attracting BEC are reliable, as the two distinct models produce
them in a nearly identical form. However, a difference between the models is
found too, as strongly pinned (very narrow) discrete solitons, which were
previously found in model 1, are not generated by model 2 -- in fact, in
agreement with the continual 1D NPSE, which does not have such solutions
either. In that respect, the newly derived model provides for a more
accurate approximation for the trapped BEC.
\end{abstract}

\maketitle

\textbf{The dynamics of a dilute quantum gas which forms the Bose-Einstein
condensate (BEC) is very accurately described by the three-dimensional
Gross-Pitaevskii equation (3D GPE). This equation treats effects of
collisions between atoms in the condensate in the mean-field approximation.
In experimentally relevant settings, the BEC\ is always confined by a
trapping potential. In many cases, the trap is designed to have the
``cigar-shaped" form, allowing an effective reduction of the dimension from
3 to 1. In turn, the 1D dynamics of the trapped condensate may be controlled
by means of an additional periodic potential, induced by an optical lattice
(OL), which acts along the axis of the ``cigar". If the OL potential is
sufficiently strong, the eventual dynamical model reduces to a 1D discrete
equation. In both the continual and discrete versions of the 1D description,
a crucially important feature is the form of the nonlinearity in the
respective equations. In the limit of low density, the nonlinearity is cubic
-- the same as in the underlying 3D GPE. In the general case, a consistent
derivation, which starts from the cubic nonlinearity in 3D, leads to 1D
equations with a nonpolynomial nonlinearity, the respective model being
called the ``nonpolynomial Schr\"{o}dinger equation" (NPSE). The discrete
limit of the latter equation, corresponding to the action of the strong
axial OL potential, was derived and investigated recently. An essential
asset of both versions of the NPSE, continual and discrete ones, is that
they predict the onset of the collapse (formation of a singularity in the
condensate with attraction between atoms) in the framework of the 1D
description, thus complying with the fundamental property of the BEC which
was predicted by the underlying 3D GPE and observed experimentally. However,
in the case when the OL potential is very strong, an alternative way to
derive the 1D discrete model may start with the discretization of the 3D
GPE, followed by the reduction of the dimension in the cigar-shaped trap. In
this work, we report a new discrete model (``model 2") derived in this way,
which seems very different from the previously known discrete 1D NPSE (which
we call ``model 1"). In particular, while model 1 amounts to a single
discrete equation for the 1D complex wave function, model 2 incorporates an
additional equation for the transverse width. Nevertheless, numerical
analysis performed in the present work shows remarkably similar behavior of
fundamental localized modes, in the form of unstaggered discrete solitons,
in both systems. The similarity pertains to the existence region for the
solitons and their stability limits. Importantly, both models admit the
collapse, and produce similar predictions for the collapse threshold. Thus,
basic properties of discrete solitons found in the model of the strongly
trapped BEC are trustworthy, as they are reproduced independently by the two
very different models. Nevertheless, a difference between the two models is
also found: very narrow discrete solitons, which exist in model 1, are
absent in model 2. In fact, the continual 1D NPSE does not give rise to such
extremely narrow solutions either, thus indicating that the newly derived
model, although having a more complex mathematical form, eventually provides
a more accurate approximation.}

\section{Introduction}

The dynamics of Bose-Einstein condensates (BECs) made of dilute
ultracold gases of bosonic atoms obeys the 3D Gross-Pitaevskii
equation (GPE) in the mean-field approximation \cite{BEC}. An
effective 1D equation can be derived from the 3D\ GPE to describe
the BEC dynamics in prolate ("cigar-shaped") traps
\cite{PerezGarcia98}-\cite{Napoli2}. In the simplest case, which
corresponds to a sufficiently low BEC density, the reduced 1D
equation amounts to the cubic nonlinear Schr\"{o}dinger equation
(NLSE) \cite{Panos}. A significant restriction in the use of the
1D cubic NLSE in this context is its failure to predict the onset
of the collapse of localized modes, which was theoretically
predicted in 3D models and experimentally observed in the
self-attractive BEC \cite{Strecker02,Cornish}. This problem can be
resolved
by the more accurate reduction of the 3D GPE to the 1D nonlinear Schr\"{o}%
dinger equation with a \textit{nonpolynomial} nonlinearity (NPSE) \cite%
{sala1}, \cite{salan}, without imposing the constraint of a very
low density. The 1D NPSE model with self-attractive nonlinearity
enables the description of the collapse dynamics, and yields
results which are accurately reproduced by direct simulations of
the underlying 3D GPE \cite{Canary}. An intermediate
approximation, which can be obtained from the expansion of the
NPSE, is represented by the 1D NLSE with the self-focusing cubic
and quintic terms.
This approximation may be sufficient for some particular purposes \cite%
{Shlyap02,Brand06,Lev}.

On the other hand, the BEC trapped in a very deep optical lattice (OL) can
be well described by the corresponding discrete equations. In particular,
discrete forms of the 1D GPE with the cubic nonlinearity \cite%
{DNLS-BEC,gpe,DNLS-BEC-review} and 1D NPSE \cite{luca,nash} have been
studied in detail. Basic features of the 1D continual models describing the
BEC trapped in a deep OL find their counterparts in the discrete models, a
significant one being the ability of the discrete 1D NPSE to describe the
onset of the collapse predicted by the corresponding continual equation \cite%
{luca,nash}.

Thus, the previously explored 1D discrete BEC models relied upon
the discretization after the reduction of the dimension from $3$
to $1$ within the framework of the continual equations. However,
an alternative approach is possible too, in the situation when the
OL is stronger than the cigar-shaped potential: one should first
discretize the 3D GPE in the axial direction, and then reduce the
dimension to $1$ \cite{salan}, opposite to the previously
developed derivation \cite{sala1}. The purpose of the present work
is to derive the discrete model in this way, which, to the best of
our knowledge, has not been done before. Naturally, the two
different routes lead to quite different discrete systems. In
particular, the one derived in this work involves two sets of
discrete variables (the wave function and transverse width),
rather than the single set on which previously studied models were
based. Our purpose is not only to derive the alternative system,
but also to study generic properties of fundamental unstaggered
solitons in it, and compare the results with those obtained in the
earlier known discrete NPSE. In fact, we will conclude that,
despite a very different form of the new model, it gives rise to
results quite similar to those produced by the discrete NPSE.
Thus, despite significant differences in the form of the
``competing" models, the physical predictions for the discrete BEC
solitons are essentially the same, which attests to the
reliability of these results.

The paper is structured as follows. The new model is derived in section 2.
Basic results for the existence, stability and dynamical behavior of
fundamental discrete solitons in it are reported and compared to their
counterparts predicted by the usual 1D discrete NPSE in section 3. The paper
is concluded by section 4.

\section{The derivation of the discrete one-dimensional systems}

The starting point of the consideration is the 3D GPE, which governs the
evolution of macroscopic wave function $\psi (\mathbf{r},t)$ of the dilute
BEC near zero temperature \cite{BEC}. This equation can be derived from the
following action functional \cite{sala1},
\begin{eqnarray}
S &=&\int \,dtd\mathbf{r}\ \psi ^{\ast }(\mathbf{r},t)\left( ih\kern-0.6em%
\char"16\kern-0.1em\frac{\partial }{\partial t}+\frac{h\kern-0.6em\char"16%
\kern-0.1em^{2}}{2m}(\nabla _{\perp }^{2}+\frac{\partial ^{2}}{\partial z^{2}%
})-V_{0}\cos {(2kz)}\right.   \notag \\
&&-\left. \frac{m\omega _{\bot }^{2}}{2}(x^{2}+y^{2})-\frac{1}{2}\gamma
(N-1)|\psi |^{2}\right) \psi (\mathbf{r},t),  \label{action}
\end{eqnarray}%
where $\gamma =4\pi h\kern-0.6em\char"16\kern-0.1em^{2}a_{s}/m$ is the
strength of the interaction between bosons, $a_{s}$ the $s$-wave scattering
length, $N$ the number of bosonic atoms, $m$ the atomic mass, and $\omega
_{\bot }$ the frequency of the transverse harmonic confinement. The
corresponding 3D GPE is written as
\begin{equation}
ih\kern-0.6em\char"16\kern-0.1em\frac{\partial }{\partial t}\psi =\left[ -%
\frac{h\kern-0.6em\char"16\kern-0.1em^{2}}{2m}\nabla ^{2}+V_{0}\cos {(2kz)}+%
\frac{m\omega _{\bot }^{2}}{2}(x^{2}+y^{2})+\gamma (N-1)|\psi |^{2}\right]
\psi \;.  \label{gpe}
\end{equation}%
Expressing the length in units of $a_{\perp }$, time in units of $\omega
_{\perp }^{-1}$ and energy in units of $h\kern-0.6em\char"16\kern%
-0.1em\omega _{\perp }$, the 3D GPE is cast in the scaled form,
\begin{equation}
i\frac{\partial }{\partial t}\psi =\left[ -\frac{1}{2}(\nabla _{\perp }^{2}+%
\frac{\partial ^{2}}{\partial z^{2}})+V_{0}\cos {(2kz)}+\frac{1}{2}%
(x^{2}+y^{2})+2\pi \Gamma |\psi |^{2}\right] \psi ,  \label{ngpe}
\end{equation}%
where $\Gamma =2(N-1)a_{s}/a_{\perp }$ is the effective strength of the
self-interaction, and $a_{\perp }=\sqrt{\hbar /(m\omega _{\bot })}$ is the
characteristic length of the transverse confinement. Notice that $a_{s}<0$,
i.e., $\Gamma <0$ in Eq. (\ref{ngpe}), corresponds to the attraction between
bosons.

Below, we consider the derivation of two alternative forms of the 1D
discrete approximation. The previously known ``model 1" is derived starting
with reduction of the 3D GPE to the 1D NPSE, which is then discretized \cite%
{luca,nash}. Here, we briefly recapitulate this route of the derivation, for
the purpose of the comparison with novel ``model 2", which is obtained from
the 3D GPE by \emph{first} discretizing it, and \emph{then} reducing the
resulting system to the 1D form.

\subsection{Model 1: the reduction of the dimension followed by the
discretization}

The derivation of model 1 starts with the dimensional reduction, which is
performed by the minimization of action functional (\ref{action}), choosing
the wave function in the form of%
\begin{equation}
\psi (\mathbf{r},t)=\frac{\exp \left[ -(x^{2}+y^{2})/(2\sigma ^{2}(z,t))%
\right] }{\sqrt{\pi }\sigma (z,t)}f(z,t),  \label{trial}
\end{equation}%
where real $\sigma $ is the local width of the transverse confinement, and
complex 1D (axial) wave function $f$ is normalized,
\begin{equation}
\int_{-\infty }^{+\infty }\left\vert f(z)\right\vert ^{2}dz=1  \label{=1}
\end{equation}%
\cite{sala1}. Inserting trial wave function (\ref{trial}) in Eq. (\ref%
{action}) and performing the integration over $x$ and $y$, the action can be
written as
\begin{equation}
S=\int \,dt\int_{-\infty }^{+\infty }dz~f^{\ast }\left( i\frac{\partial }{%
\partial t}+\frac{1}{2}\frac{\partial ^{2}}{\partial z^{2}}-V_{0}\cos {(2qz)}%
+\frac{\Gamma }{\sigma ^{2}}|\psi |^{2}-\frac{1}{2}(\frac{1}{\sigma ^{2}}%
+\sigma ^{2})\right) f\;,  \label{action1}
\end{equation}%
in the approximation which neglects $\partial \sigma /\partial z$ \cite%
{sala1}. The Euler-Lagrange equations derived from Eq. (\ref{action1}) by
varying with respect to $f^{\ast }$ and $\sigma $,
\begin{eqnarray}
i\frac{\partial f}{\partial t} &=&\left[ -\frac{1}{2}\frac{\partial ^{2}}{%
\partial z^{2}}+V_{0}\cos {(2qz)}+\frac{\Gamma }{\sigma ^{2}}|f|^{2}+\frac{1%
}{2}\left( \frac{1}{\sigma ^{2}}+\sigma ^{2}\right) \right] f,  \label{amp}
\\
\sigma ^{4} &=&1+\Gamma |f|^{2},  \label{shi}
\end{eqnarray}%
may be combined into the equation (NPSE) derived in Ref. \cite{sala1},
namely,
\begin{equation}
i\frac{\partial f}{\partial t}=\left[ -{\frac{1}{2}}{\frac{\partial ^{2}}{%
\partial z^{2}}}+V_{0}\cos {(2qz)}+\frac{1+(3/2)\Gamma |f|^{2}}{\sqrt{%
1+\Gamma |f|^{2}}}\right] f\;.  \label{npse1}
\end{equation}

Assuming that OL potential in this equation is strong enough, one can
further derive the discrete version of the 1D NPSE \cite{luca,nash}. To this
end, the continual wave function is approximated by a superposition of
orthonormal modes $W_{n}$ (such as Wannier functions), which are tightly
confined in a vicinity of local potential minima, $z_{n}=\pi n/q$, with
integer $n$:
\begin{equation}
f(z,t)=\sum_{n}f_{n}(t)W_{n}(z),  \label{quasi-Wannier}
\end{equation}%
$f_{n}$ being the respective complex amplitudes. This decomposition is made
unique by imposing a condition that the largest value of each local function
$W_{n}$ is $1$ \cite{luca}. Next, one may insert this ansatz in Eq. (\ref%
{amp}), multiply the resulting equation by the complex conjugate of the
local mode, and integrate over $z$. The so derived 1D discrete NPSE,
together with the discrete version of equation (\ref{shi}), take the form of%
\begin{eqnarray}
i\frac{\partial }{\partial t}f_{n} &=&\left[ \frac{1}{2}\left( \frac{1}{%
\sigma _{n}^{2}}+\sigma _{n}^{2}\right) +\epsilon \right]
f_{n}-C(f_{n+1}+f_{n-1})+\frac{g}{\sigma _{n}^{2}}|f_{n}|^{2}f_{n},
\label{eq3r2f} \\
\sigma _{n}^{4} &=&1+g|f_{n}|^{2},  \label{eq3r2s}
\end{eqnarray}%
where the local norms and parameters $\epsilon ,C$ and $g$ are
\begin{eqnarray}
\epsilon  &\equiv &\int W_{n}^{\ast }(z)\left[ -\frac{1}{2}\frac{\partial
^{2}}{\partial z^{2}}+V_{0}\cos {(2kz)}\right] W_{n}(z)\,dz,  \notag \\
C &\equiv &-\int W_{n+1}^{\ast }(z)\left[ -\frac{1}{2}\frac{\partial ^{2}}{%
\partial z^{2}}+V_{0}\cos {(2kz)}\right] W_{n}(z)\,dz,  \notag \\
g &\equiv &\Gamma \int |W_{n}(z)|^{4}\,dz.  \label{J}
\end{eqnarray}%
In the tight-binding approximation, $C$ is positive definite.

In what follows below, the discrete system based on Eqs. (\ref{eq3r2f}) and (%
\ref{eq3r2s}) is referred to as ``model 1". These equations correspond to
the Lagrangian,
\begin{equation}
L_{\mathrm{eff}}=\sum_{n}\left\{ f_{n}^{\ast }\left[ i\frac{\partial }{%
\partial t}-\frac{1}{2}\left( \frac{1}{\sigma _{n}^{2}}+\sigma
_{n}^{2}\right) -\epsilon \right] f_{n}+Cf_{n}^{\ast }(f_{n+1}+f_{n-1})-%
\frac{g}{2\sigma _{n}^{2}}|f_{n}|^{4}\right\} .
\end{equation}%
Model 1 conserves two dynamical invariants, \textit{viz}., the norm (alias
``power") and Hamiltonian,%
\begin{equation}
\mathcal{N}=\sum_{n}~|f_{n}|^{2},~\mathcal{H}=\sum_{n}{\left(
C|f_{n}-f_{n+1}|^{2}+\sqrt{1-g|f_{n}|^{2}}|f_{n}|^{2}\right) }.  \label{NH}
\end{equation}

\subsection{Model 2: the discretization followed by the reduction of the
dimension}

The derivation of the novel system starts with the direct discretization of
the 3D GPE, i.e., Eq. (\ref{gpe}), by adopting the following ansatz for the
wave function,
\begin{equation}
\psi (x,y,z,t)=\sum_{n}\phi _{n}(x,y)W_{n}(z),  \label{wan2}
\end{equation}%
where $W_{n}(z)$ is the same set of local modes as in Eq. (\ref%
{quasi-Wannier}). We insert this ansatz in Eq. (\ref{gpe}), multiply the
resulting equation by $W_{n}^{\ast }$, and integrate the result over $z$.
This leads to the semi-discrete equation,
\begin{equation}
i\frac{\partial }{\partial t}\phi _{n}=\left[ -\frac{1}{2}\nabla _{\perp
}^{2}+\frac{1}{2}(x^{2}+y^{2})+\epsilon \right] \phi _{n}-C(\phi _{n+1}+\phi
_{n-1})+2\pi g|\phi _{n}|^{2}\phi _{n},  \label{eq3r2}
\end{equation}%
where $\nabla _{\perp }^{2}$ acts on coordinates $x$ and $y$, and parameters
$\epsilon ,C$ and $g$ are defined as per Eqs. (\ref{J}).

Equation (\ref{eq3r2}) can be derived from the corresponding Lagrangian,
\begin{equation}
L=\int_{-\infty }^{+\infty }dx\int_{-\infty }^{+\infty }dy\sum_{n}\left[
\phi _{n}^{\ast }\left[ i\frac{\partial }{\partial t}+\frac{1}{2}\nabla
_{\perp }^{2}-\frac{1}{2}\left( x^{2}+y^{2}\right) -\epsilon \right] \phi
_{n}+C\phi _{n}^{\ast }(\phi _{n+1}+\phi _{n-1})-\pi g|\phi _{n}|^{4}\right]
.  \label{eq7r2}
\end{equation}%
The further simplification is performed by substituting ansatz
\begin{equation}
\phi _{n}(x,y,t)=\frac{1}{\sqrt{\pi }\sigma _{n}(t)}\exp \left( {-\frac{%
x^{2}+y^{2}}{2\sigma _{n}^{2}(t)}}\right) f_{n}(t)  \label{ansatz2}
\end{equation}%
into Lagrangian (\ref{eq7r2}) and integrating over the $\left( x,y\right) $
plane, which eventually leads to an effective Lagrangian for 1D discrete
fields $f_{n}(t)$ and $\sigma _{n}(t)$,
\begin{gather}
L=\sum_{n}\left\{ f_{n}^{\ast }\left[ i\frac{\partial }{\partial t}-\frac{1}{%
2}\left( \frac{1}{\sigma _{n}^{2}}+\sigma _{n}^{2}\right) -\epsilon \right]
f_{n}\right.   \notag \\
\left. +2C\frac{\sigma _{n}\sigma _{n+1}}{\sigma _{n}^{2}+\sigma _{n+1}^{2}}%
(f_{n}^{\ast }f_{n+1}+f_{n-1}^{\ast }f_{n})-\frac{g}{2\sigma _{n}^{2}}%
|f_{n}|^{4}\right\} .  \label{eq14r2}
\end{gather}%
The Euler-Lagrange equations derived from Eq. (\ref{eq14r2}) by varying the
Lagrangian with respect to $f_{n}^{\ast }$ and $\sigma _{n}$ are
\begin{equation}
i\frac{\partial }{\partial t}f_{n}=\left[ \frac{1}{2}\left( \frac{1}{\sigma
_{n}^{2}}+\sigma _{n}^{2}\right) +\epsilon \right] f_{n}+\frac{g}{\sigma
_{n}^{2}}|f_{n}|^{2}f_{n}-C\left( \frac{2\sigma _{n}\sigma _{n+1}}{\sigma
_{n}^{2}+\sigma _{n+1}^{2}}f_{n+1}+\frac{2\sigma _{n}\sigma _{n-1}}{\sigma
_{n}^{2}+\sigma _{n-1}^{2}}f_{n-1}\right) ,  \label{ampeq}
\end{equation}%
\begin{gather}
|f_{n}|^{2}\frac{1+g|f_{n}|^{2}-\sigma _{n}^{4}}{\sigma _{n}^{3}}%
+2C(f_{n+1}^{\ast }f_{n}+f_{n+1}f_{n}^{\ast })\frac{\sigma _{n+1}(\sigma
_{n+1}^{2}-\sigma _{n}^{2})}{(\sigma _{n+1}^{2}+\sigma _{n}^{2})^{2}}  \notag
\\
+2C(f_{n-1}^{\ast }f_{n}+f_{n-1}f_{n}^{\ast })\frac{\sigma _{n-1}(\sigma
_{n-1}^{2}-\sigma _{n}^{2})}{(\sigma _{n-1}^{2}+\sigma _{n}^{2})^{2}}=0.
\label{shieq}
\end{gather}%
Equations (\ref{ampeq}) and (\ref{shieq}) conserve the respective norm and
Hamiltonian, which are [cf. Eq. (\ref{NH})]%
\begin{eqnarray}
\mathcal{N} &=&\sum_{n}|f_{n}|^{2},  \notag \\
\mathcal{H} &=&\sum_{n}\left\{ f_{n}^{\ast }\left[ \frac{1}{2}\left( \frac{1%
}{\sigma _{n}^{2}}+\sigma _{n}^{2}\right) +\epsilon \right] f_{n}-2C\frac{%
\sigma _{n}\sigma _{n+1}}{\sigma _{n}^{2}+\sigma _{n+1}^{2}}(f_{n}^{\ast
}f_{n+1}+f_{n-1}^{\ast }f_{n})+\frac{g}{2\sigma _{n}^{2}}|f_{n}|^{4}\right\}
.  \label{NHa}
\end{eqnarray}

It is worthy to note that Eq. (\ref{eq3r2s}) in model 1 involves only $%
\sigma _{n}$ (without coupling to $\sigma _{n\pm 1}$), because the
corresponding spatial derivative, $\partial \sigma /\partial z$, was
neglected in the underlying expression (\ref{action1}). On the contrary to
that, Eq. (\ref{shieq}) in model 2 couples $\sigma _{n}$ to $\sigma _{n\pm 1}
$, as this coupling is implicitly retained by expressions (\ref{eq7r2}) and (%
\ref{ansatz2}). For that reason, $\sigma _{n}^{2}$ cannot be
explicitly found from Eq. (\ref{shieq}), hence model 2, unlike
model 1, cannot be reduced to a single equation for the discrete
wave function. Nevertheless, model 2 is amenable to a consistent
numerical analysis, see below. The two models can be made formally
equivalent only if transverse widths $\sigma _{n} $ are postulated
to be constant ($t$- and $n$-independent), in which case either
system reduces to the standard discrete 1D NLSE. Finally, it is
worthy to mention that both models, i.e., Eqs. (\ref{eq3r2f}),
(\ref{eq3r2s}), on the one side and Eq. (\ref{ampeq}) on the
other, take identical limit forms in the anticontinuum limit, $C =
0$.

\section{Fundamental bright solitons}

\subsection{The existence of fundamental solitons}

In this section we present families of fundamental bright-soliton solutions
to Eqs. (\ref{ampeq}) and (\ref{shieq}), identify their stability and
compare them to the corresponding soliton families in model 1 \cite%
{luca,nash}. If the nonlinearity is attractive ($g<0$), the stationary
solutions of models 1 and 2 with chemical potential $\mu $ are sought for by
the substitution into Eqs. (\ref{eq3r2f}), (\ref{eq3r2s}) and (\ref{ampeq}),
respectively, of
\begin{equation}
f_{n}(t)=|g|^{-1/2}u_{n}\exp \left( -i\mu t\right) .  \label{fu}
\end{equation}%
In doing so, we assume that $\epsilon =2C$ was fixed by means of an obvious
additional shift of the chemical potential. Then, real discrete functions $%
u_{n}$ obey stationary equations, which are%
\begin{equation}
\mu u_{n}=-{C}\left( u_{n+1}+u_{n-1}-2u_{n}\right) +\frac{1-(3/2)u_{n}^{2}}{%
\sqrt{1-u_{n}^{2}}}u_{n}  \label{st1}
\end{equation}%
in model 1, and
\begin{eqnarray}
\mu u_{n} &=&-\left[ \frac{1}{2}\left( \frac{1}{\sigma _{n}^{2}}+\sigma
_{n}^{2}\right) +2C\right] u_{n}+\frac{g}{\sigma _{n}^{2}}|u_{n}|^{2}u_{n}
\notag \\
&&-C\left( \frac{2\sigma _{n}\sigma _{n+1}}{\sigma _{n}^{2}+\sigma _{n+1}^{2}%
}u_{n+1}+\frac{2\sigma _{n}\sigma _{n-1}}{\sigma _{n}^{2}+\sigma _{n-1}^{2}}%
u_{n-1}\right)   \label{st2}
\end{eqnarray}%
in model 2. In the latter case, widths $\sigma _{n}$ are to be found from
Eq. (\ref{shieq}), which takes the form of%
\begin{eqnarray}
|u_{n}|^{2}\frac{1+g|u_{n}|^{2}-\sigma _{n}^{4}}{\sigma _{n}^{3}}
&+&2C(u_{n+1}^{\ast }u_{n}+u_{n+1}u_{n}^{\ast })\frac{\sigma _{n+1}(\sigma
_{n+1}^{2}-\sigma _{n}^{2})}{(\sigma _{n+1}^{2}+\sigma _{n}^{2})^{2}}  \notag
\\
&&+2C(u_{n-1}^{\ast }u_{n}+u_{n-1}u_{n}^{\ast })\frac{\sigma _{n-1}(\sigma
_{n-1}^{2}-\sigma _{n}^{2})}{(\sigma _{n-1}^{2}+\sigma _{n}^{2})^{2}}=0.
\label{stsh2}
\end{eqnarray}

Stationary equations (\ref{st1}) and (\ref{st2}) were solved numerically,
using an algorithm based on the modified Powell minimization method \cite%
{luca,gpe}. The initial ansatz used to construct on-site and
inter-site-centered discrete solitons in model 1 was, respectively, $\left\{
u_{n}^{(0)}\right\} =(...,\,0,\,A,\,0,\,...)$ and $(...,\,0,\,A,\,A,\,0,%
\,...)$, where $A$ is a real constant obtained from Eq. (\ref{st1}) in the
corresponding approximation. These soliton solutions are then used as an
initial ansatz to generate discrete solitons in model 2. Results reported
below were obtained in the lattice composed of $101$ or $100$ sites, for the
on-site and inter-site configurations, respectively. It was checked that the
results do not alter if a larger lattice had been used.

\begin{figure}[tbp]
\center\includegraphics [width=12.7cm]{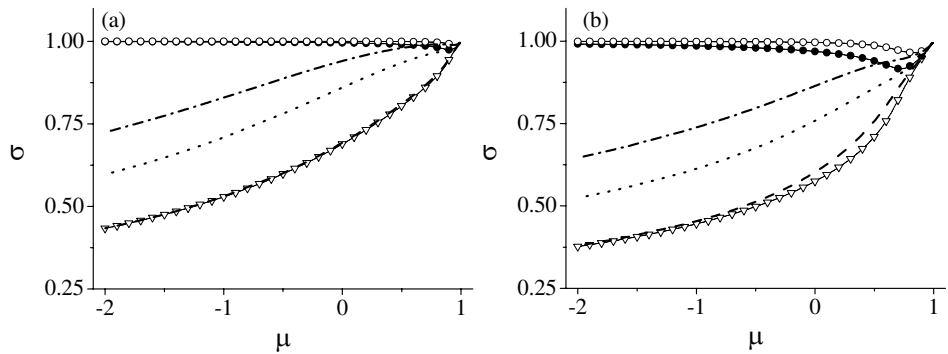}
\caption{Values of the transverse width, $\protect\sigma _{n}$, as functions
of the chemical potential, $\protect\mu $, for fundamental on-site solitons
in model 1 (the ordinary model) are shown by curves which are marked by
triangles for the central site, full circles -- for the first neighbors, and
empty circles -- for the second neighbors to the central site. In model 2
(the new system) the values of $\protect\sigma _{n}$ for fundamental on-site
solitons are shown by dashed lines for the central site, dotted lines -- the
first neighbors, and dashed-dotted lines -- the second neighbors to the
central site. The inter-site coupling constant is $C=0.2$ (a) and $C=0.8$
(b). }
\label{fig1a}
\end{figure}

\begin{figure}[tbp]
\center\includegraphics [width=12.7cm]{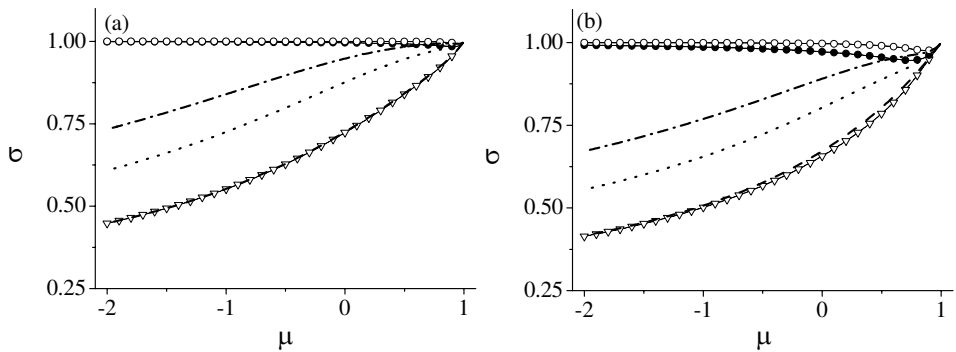}
\caption{The same as in Fig. \protect\ref{fig1a}, but for families of
inter-site solitons found in both models 1 and 2.}
\label{fig1b}
\end{figure}

We start the presentation of the results by plotting, in Figs. \ref{fig1a}
and \ref{fig1b}, transverse widths $\sigma _{n}$ versus the chemical
potential for stationary unstaggered solitons\ of the on-site and inter-site
types, respectively, found in both models 1 and 2 [recall that the vanishing
field corresponds to $\sigma _{n}=1$, see Eqs. (\ref{eq3r2s}) and (\ref%
{shieq})]. It is evident that these characteristics of the soliton families
are close for both models, with some difference observed in central parts of
the solitons.

\begin{figure}[tbp]
\center\includegraphics [width=12cm]{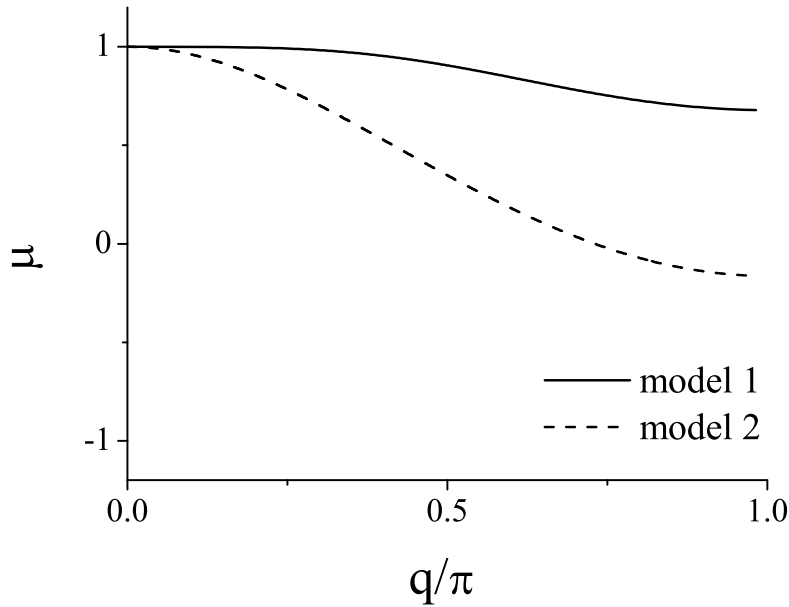}
\caption{The CW solutions are unstable to modulational perturbations in
regions below curves $\protect\mu (q)$, which are determined by condition $%
\Omega ^{2}>0$, see Eq. (\protect\ref{dispersion}). The solid and dashed
curves correspond to models 1 and 2, respectively. The figure pertains to $%
C=0.8$.}
\label{figd}
\end{figure}

To outline the entire existence region for the fundamental solitons, we
followed the usual approach, identifying it as the region where CW
(continuous-wave) solutions are modulationally unstable, hence the existence
of bright solitons should be expected. The CW solutions can be easily found
from stationary equations (\ref{st1}) and (\ref{st2}) in the form of%
\begin{equation}
u_{n}=Ue^{-i\mu t},~\mu _{\mathrm{CW}}=\left[ 1+\left( 3/2\right) gU^{2}%
\right] /\sqrt{(1+gU^{2})},~\sigma _{n}^{2}=1+gU^{2},
\end{equation}%
which shows that in the case of the attractive interaction, $g=-1$, the
amplitude of the CW solution is subject to constraint $U<1$, and the
respective chemical potential takes values $\mu _{\mathrm{CW}}<1$. Then,
straightforward calculations yield a dispersion relation for frequency $%
\Omega $ and wavenumber $q$ of small modulational perturbations around the
CW solution:
\begin{equation}
\Omega ^{2}=-2A\left[ 2A+\frac{2gU^{2}}{\sqrt{1+gU^{2}}}-\frac{g^{2}U^{4}}{%
2(1+gU^{2})\left( \sqrt{1+gU^{2}}+D\right) }\right] ,  \label{dispersion}
\end{equation}%
where $A\equiv 2C\sin ^{2}\left( q/2\right) $, and $D=A$ in model 2, $D=0$
in model 1. Analysis of this dispersion relation demonstrates that, in
either model, the CW solution is unstable for $\mu <1$ against long-wave
perturbations (for small values of $q$), see Fig. (\ref{figd}).

In the case of the repulsive contact interactions, $g>0$, one may expect the
existence of \textit{staggered} discrete solitons \cite{DNLS-BEC-review}.
The analysis of the modulational instability of the respective staggered CW
solutions (not given here in detail) shows that, for both models, the
staggered CW states are indeed unstable in their entire existence region ($%
\mu >1$), which indicates that staggered solitons may exist at $\mu >1$ in
either model. The further study of model 1 indicates that it supports
strongly localized (tightly pinned) staggered solitons of the
fundamental type (as reported before, see Figs. 20-23 in Ref. \cite{luca}). Model 2 reproduces only low-amplitude
tightly pinned staggered modes near the lower boundary of the existence region.
Those solutions are not displayed here, as the objective of the work is to focus
on unstaggered solitons in the case of the self-attraction, when the
difference of both models, 1 and 2, from the ordinary discrete cubic NLSE is
most essential.

\subsection{The norm and free energy of fundamental unstaggered solitons}

\begin{figure}[tbp]
\center\includegraphics [width=12.7cm]{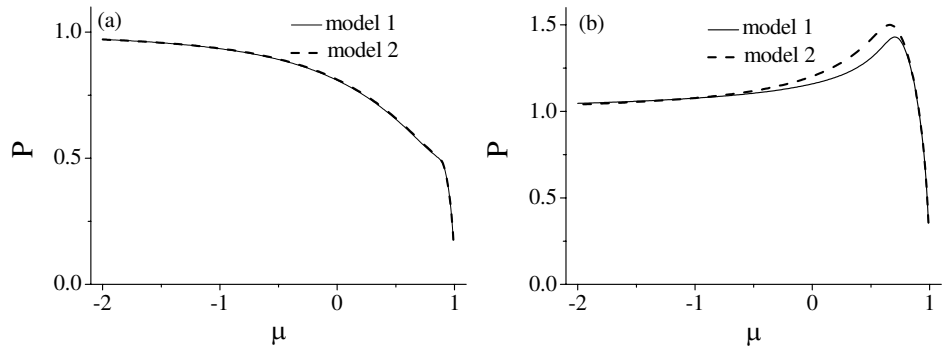}
\caption{The norm ("power") $P$ versus chemical potential $\protect\mu $
for on-site fundamental solitons, in both models 1
and 2. The coupling constants are $C=0.2$ (a) and $C=0.8$ (b), which correspond to
strongly discrete systems and quasi-continual systems, respectively.}
\label{fig2a}
\end{figure}

\begin{figure}[tbp]
\center\includegraphics [width=12.7cm]{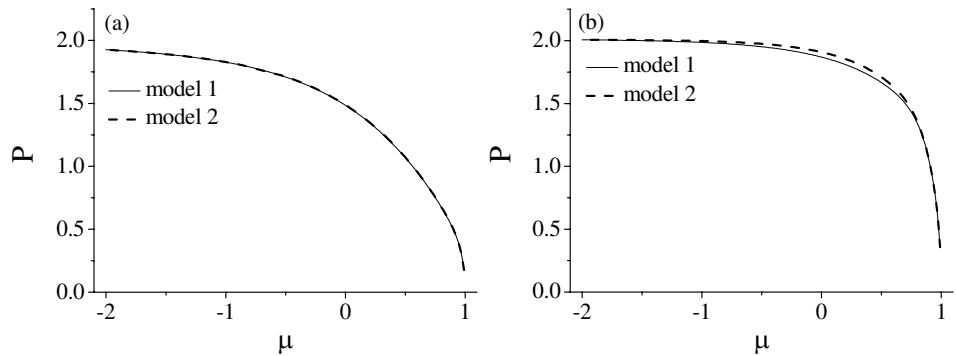}
\caption{The norm ("power") $P$ versus chemical potential $\protect\mu $
for inter-site fundamental solitons, in models 1
and 2. The coupling constants are $C=0.2$ (a) and $C=0.8$ (b).
}
\label{fig2b}
\end{figure}

\begin{figure}[tbp]
\center\includegraphics [width=12.7cm]{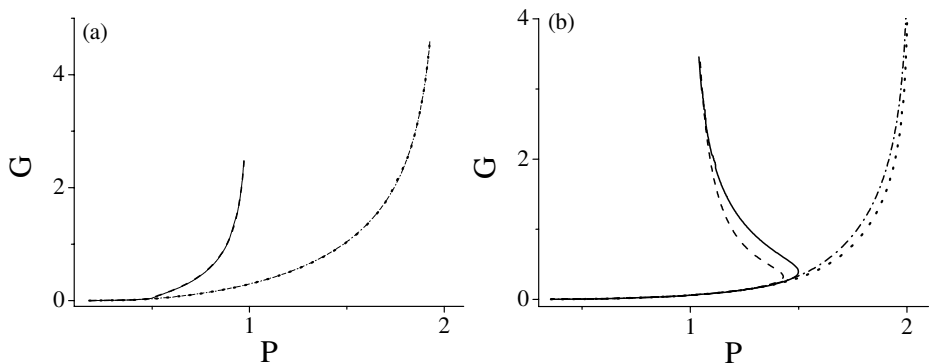}
\caption{Free energy $G$ versus norm $P$ for on-site and inter-site
fundamental solitons, in both models 1 and 2, for $C=0.2$ (a) and $C=0.8$
(b). Solid and dashed lines correspond to the on-site solitons in models 2
and 1, respectively. Dotted and dashed-dotted lines represent, respectively,
inter-site solitons in models 2 and 1.}
\label{fig3}
\end{figure}

The power (norm) of unstaggered solitons in the two models, $%
P=\sum_{n=1}^{N}|u_{n}|^{2}$, and their free energy, $G\equiv H-\mu P$,
where $H$ is obtained by inserting stationary solution (\ref{fu}) into
expressions (\ref{shieq}) or (\ref{NHa}), are compared in Figs. (\ref{fig2a}%
), (\ref{fig2b}) and (\ref{fig3}) for two values of the intersite-coupling
constant, $C=0.2$ and $C=0.8$. These values, which were used in Figs. \ref%
{fig1a} and \ref{fig2b}, actually correspond, severally, to limit cases of
strongly discrete and quasi-continual systems. It is seen that the norms and
free energies of the soliton families in both models feature similar
behaviors: the same sign of the slope, $dP/d\mu $, for both the on-site and
inter-site modes, the equality of the on-site and inter-site free energies
in the region of a small power and small $|\mu |$ (wide solitons with a
small amplitude). Quantitative differences between the norm and free energy
between the two models are larger for higher values of $C$, i.e., closer to
the continuum limit. Some differences also appear in the region of large
norms, which corresponds to very narrow solitons, where the $P(\mu )$ curves
show a trend to saturation. In particular, in that region amplitude $%
u_{n}^{2}$ of the narrow solitons in models 2 may exceed $1$, which is the
absolute upper limit (collapse threshold) in model 1, as follows from Eq. (%
\ref{shi}). In fact, numerical results indicate at the existence of a
similar limit in model 2, although it does not explicitly follow from Eqs. (%
\ref{ampeq}) and (\ref{shieq}). Finding exact values for this limit is
complicated by difficulties in obtaining accurate numerical results for very
narrow (strongly pinned) solitons in model 2.

\subsection{Dynamical considerations}

The sufficient condition for soliton stability in the NPSE is the
spectral condition according to which the corresponding
eigenvalues, found from linearized equations for small
perturbations, must not have a positive real part \cite{gpe,
stability}. Although the Vakhitov-Kolokolov (alias \textit{slope})
stability criterion, $dP/d\mu \leq 0$ \cite{VK}, is not strictly applicable
to the models with nonpolynomial nonlinearity we will also briefly
comment it.

In Figs. (\ref{fig2a}) and (\ref{fig2b}), $P(\mu )$ curves for the
fundamental-soliton families have the same sign of the slope ($dP/d\mu $) in
models 1 and 2. Therefore, in the region of small $P$ and small positive $%
\mu $, the on-site solitons may be stable in both models, according to the
VK criterion. In all other cases, the on-site solitons are expected to be
unstable. On the other hand, the VK criterion indicates the stability of
inter-site solitons in the entire existence region, in both models.

\begin{figure}[tbp]
\center\includegraphics [width=7.7cm]{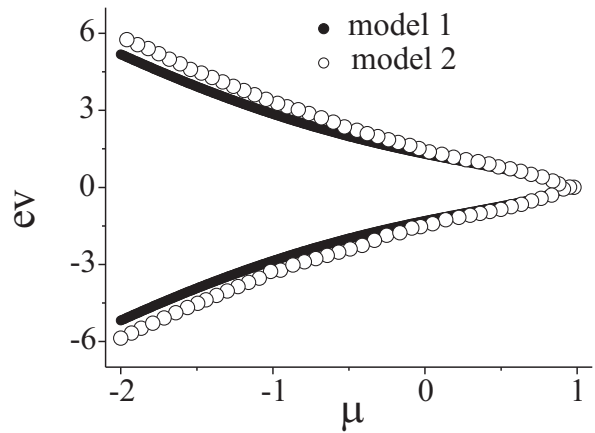}
\caption{Pure real unstable eigenvalues ("ev") for inter-site unstaggered
solitons in models 1 and 2 are shown by black and white circles,
respectively, for $C=0.2$. Note that, in this case, pure real eigenvalues
for on-site solitons have not been found. }
\label{fig4}
\end{figure}

\begin{figure}[tbp]
\center\includegraphics [width=12.7cm]{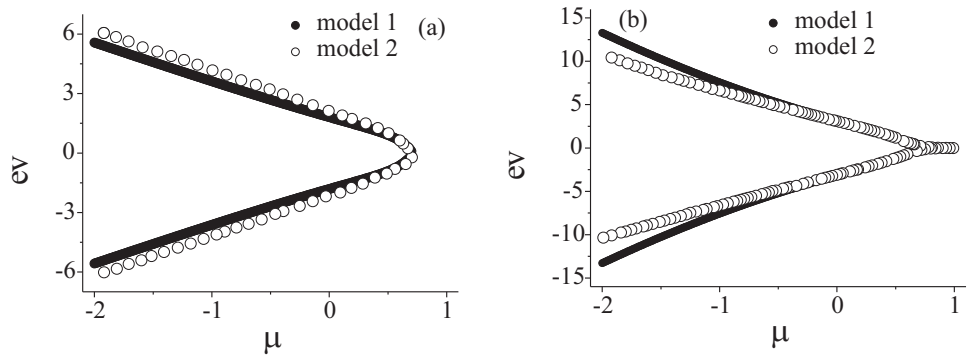}
\caption{The same as in Fig. \protect\ref{fig4}, but at $C=0.8$. In this
case, plots (a) and (b) display the eigenvalues for the on-site solitons and
inter-site solitons, respectively.}
\label{fig5}
\end{figure}

To confirm the stability of the solitons, we have checked the spectral
condition within the framework of the linear stability analysis, following
the approach elaborated in Refs. \cite{luca,nash}). The results show that
the fundamental unstaggered solitons in both models indeed have the same
stability properties, as seen in Figs. (\ref{fig4}) and (\ref{fig5}). In
particular, for on-site solitons the stability takes place only in a narrow
region in the plane of $\left( \mu ,C\right) $. This region is characterized
by very close values of the norms of the on-site and inter-site solitons,
hence a very small difference in the respective values the free energy,
which, in turn, implies a very low Peierls-Nabarro barrier \cite{luca,nash},
i.e., mobility of the solitons.

All conclusions concerning the mobility, perturbed evolution, and
development of the collapse instability are qualitatively identical in both
models. In particular, unstable inter-site modes which find on-site
counterparts with close values of the norm (or free energy), and unstable
on-site modes with close inter-site counterparts evolve into robust
breathers, with almost no loss of the norm. In the opposite case, the
instability leads to collapse of the localized modes. Our calculations show
that collapse thresholds are nearly identical in models 1 and 2. In Ref.
\cite{luca}, it was found, in the framework of model 1, that (as mentioned
above) $u_{n}^{2}$ in soliton solutions cannot exceed $1$; actually, $%
u_{n}^{2}=1$ in Eq. (\ref{st1}) is the collapse threshold, which is
determined by the singular structure of the on-site nonlinearity. A similar
trend is demonstrated by numerical results obtained in model 2.

Summarizing this section, we conclude that both discrete models, 1 and 2,
translate fundamental properties of bright solitons, known in the 1D
continual NPSE, into the discrete setting in essentially the same way (for
the discrete model of type 1, the correspondence with the continual equation
was demonstrated in recent work \cite{nascont}). The most significant among
these properties is the possibility to describe the onset of the collapse in
the framework of the 1D geometry.

\section{Conclusion}

The mean-field dynamics of BEC confined in a combination of a cigar-shaped
trap and axial OL (optical-lattice) potential, is well approximated by the
1D NPSE (nonpolynomial Schr\"{o}dinger equation), in both continual \cite%
{sala1,nascont} and discrete settings \cite{luca,nash}. In
particular, both versions of the NPSE admit a possibility to
describe the onset of the collapse of the localized modes in the
framework of the 1D approximation, thus making this approximation
compliant with the well-known fundamental property of the
self-attractive BEC confined in tight traps. The objective of the
present work was to verify the relevance of the description based
on the discrete limit of the NPSE-type equation, by comparing two
alternative versions of this approximation: the previously derived
one ("model 1"), which is based on the reduction of the underlying
3D GPE to 1D continual NPSE, followed by its discretization, and
the new "model 2", whose derivation starts with the discretization
of the 3D GPE, followed by the reduction of the dimension. The two
steps of the derivation, \textit{viz}., the dimension reduction
and discretization, are apparently non-commutative, and,
accordingly, final forms of the two models appear to be vastly
different; in particular, model 1 reduces to a single discrete
equation for the wave function, while model 2 includes an
additional equation for the transverse width. Nevertheless, the
numerical solutions for families of fundamental unstaggered bight
solitons produce \emph{very similar} results in \emph{both
models}, as concerns their existence and stability, and,
especially, the crucially important feature -- the collapse
threshold. A difference between the two models was found too, as
model 2 (the newly introduced one) does not reproduce the region
of high-amplitude strongly pinned narrow unstaggered and staggered
modes, which was predicted in model 1. Actually, that feature of
model 1 is impugnable, as it is not reproduced by the continual
NPSE with the strong OL potential, as noted in Ref. \cite{nash}.
In this respect, model 2 (the novel one), although having a more
cumbersome mathematical form, seems physically preferable.

G.G., A.M. and Lj.H. acknowledge support from the Ministry of Science,
Serbia (through project 141034). L.S. and B.A.M. appreciate a partial
support from CARIPARO Foundation, through ``Progetti di Eccellenza 2006".
The work of B.A.M. was also supported, in a part, by the German-Israel
Foundation through grant No. 149/2006.

\end{document}